\documentclass[epj]{svjour}
\usepackage{graphics}
\usepackage{axodraw}
\usepackage{amsfonts}
\usepackage{amsmath}
\usepackage{amssymb}
\usepackage{latexsym}
\usepackage{color}
\usepackage{colordvi}
\usepackage{epsfig}
\usepackage{array}
\usepackage{pifont}
\usepackage{geometry}

\newcommand{\beq}{\begin{eqnarray}}
\newcommand{\eeq}{\end{eqnarray}}

\newcommand{\be}{\begin{equation}}
\newcommand{\ee}{\end{equation}}
\newcommand{\lwrsim}{\raise0.3ex\hbox{$<$\kern-0.75em\raise-1.1ex\hbox{$\sim$}}}

\def\C2#1#2{({\cal C}_2)_{#1}^{#2}}

\def\eq#1{Eq.~(\ref{#1})}

\def\simquand#1{\mbox{\raisebox{-1.2ex}[0.ex][1.6ex]{$\widetilde{\scriptstyle #1}$}}}
\newcommand{\ghostSD}{\begin{picture}(150,25)(0,0)
\SetWidth{1.2}
\DashArrowLine(12.5,0)(37.5,0){5}
\DashArrowLine(37.5,0)(75,0){5}
\DashLine(75,0)(112.5,0){5}
\DashArrowLine(112.5,0)(137.5,0){5}
\SetWidth{1}
\Vertex(112.5,0){2}
\GlueArc(75,0)(37.5,0,90){-4}{6}
\GlueArc(75,0)(37.5,90,180){-4}{6}
\CCirc(75,0){5}{Black}{Yellow}
\CCirc(75,37.5){5}{Black}{Yellow}
\CCirc(37.5,0){5}{Black}{Yellow}
\Text(20,-10)[l]{a,k}
\Text(50,15)[l]{d,$\nu$}
\Text(100,-10)[l]{e}
\Text(100,15)[r]{f,$\mu$}
\Text(50,-10)[l]{c,q}
\Text(120,-10)[l]{b,k}
\Text(75,48)[c]{q-k}
\end{picture}}

\begin{document}
\title{Constraints on the IR behaviour of gluon and ghost propagator from Ward-Slavnov-Taylor identities}
\author{Ph.~Boucaud\inst{1}, J.P.~Leroy\inst{1}, A.~Le~Yaouanc\inst{1} 
A.Y.~Lokhov\inst{2}, J. Micheli\inst{1}, O. P\`ene\inst{1}, J.~Rodr\'iguez-Quintero\inst{3} \and 
C.~Roiesnel\inst{2}
}                     
%
%
\institute{$^a$Laboratoire de Physique Th\'eorique et Hautes
Energies; Universit\'e de Paris XI, B\^atiment 211, 91405 Orsay Cedex,
France 
\and
Centre de Physique Th\'eorique de l'Ecole Polytechnique; F91128 Palaiseau cedex, France
\and
Dpto. F\'isica Aplicada, Fac. Ciencias Experimentales; Universidad de Huelva, 21071 Huelva, Spain.
}
\date{Received: date / Revised version: date}
%
\abstract{
We consider the constraints of the Slavnov-Taylor identity of the IR behaviour of gluon and ghost 
propagators and their compatibility with solutions of the ghost Dyson-Schwinger equation and with 
the lattice picture. 
\PACS{
      {PACS-key}{discribing text of that key}   \and
      {PACS-key}{discribing text of that key}
     } 
} 
\maketitle
\section{Introduction}
\label{intro}

In ref. ~\cite{Boucaud:2005ce} we have considered the constraints on the propagator dressing 
functions which can be derived from the Ward-Slavnov-Taylor identity (WSTI) --supplemented with some 
minimal  assumptions on the analytic behaviour of the former and of the vertex form factors-- and 
we were confronted with a contradiction between them and the ones that stem from the Dyson-Schwinger 
equation. The analysis of the ghost propagator Dyson-Schwinger equation seems to indicate that only 
a non-divergent gluon can match the lattice picture for the infrared behaviour of Landau gauge Green 
functions. On the other hand, WSTI seems to require that the gluon propagator diverges while the 
ghost dressing function should be finite and non-vani\-shing. In that ref.~\cite{Boucaud:2005ce} we proposed, 
as a possible way out, that the ghost-gluon vertex function was singular (which does not contradict Taylor's 
theorem contrary to frequent claims). That hypothesis did not look very natural and 
the futher work of~\cite{Sternbeck:2005re} made it even less plausible. 

In view of the very general validity of the WSTI this situation is rather embarrassing 
and we wish to reconsider the problem. In the following, we will re-analyse the problem and 
clarify the working hypotheses to conclude either that the gluon propagator 
diverges~\footnote{Although softly enough as not to contradict the apparent 
finiteness previously stated from lattice data.} or that some of these hypoteses 
should fail.

  
\section{Notations and main hypotheses}
\label{notations}
   
We use the following notations~\cite{Boucaud:2005ce}:
\beq\label{Def} 
\left( F^{(2)}\right)^{ab}(k^2) &=& - \delta^{ab} \ \frac{F(k^2)}{k^2} 
\nonumber \\ \left( G^{(2)}_{\mu\nu}\right)^{ab}(k^2) &=& 
\delta^{ab} \ \frac{G(k^2)}{k^2}  \left( \delta_{\mu\nu}-\frac{k_\mu
k_\nu}{k^2}\right) , \nonumber \\
\eeq
where $G^{(2)}$ and $F^{(2)}$ are respectively the gluon and ghost
propagators, $G$ and $F$ are respectively the gluon and ghost dressing
functions. The ghost-gluon vertex 
$\widetilde{\Gamma}_{\mu}(p,k;q)$ ($k$ and $-p$ are the momenta of the incoming and
outgoing ghosts and $q$ the gluon momentum) is defined as follows:
\beq\label{Def2}
\Gamma_\mu^{abc}(p,k;q) &=& g_0  (-i p_\nu)  f^{abc} \widetilde{\Gamma}_{\nu\mu}(p,k;q) 
\nonumber \\
&=& i g_0 f^{abc} \ \widetilde{\Gamma}_\mu(p,k;q) \ . 
\eeq
It will be also useful to define the following scalars $H_1$ and $H_2$:
\beq\label{Def3a}
\widetilde{\Gamma}_{\mu}(-q,k;q-k) &=& q_\mu
H_1(q,k)+(q-k)_{\mu}H_2(q,k) \nonumber \\
\eeq
that, after applying the standard 
tensor decomposition~\cite{Chetyrkin:2000dq}, 
\beq
\widetilde{\Gamma}_{\nu\mu}(p,k;q)
&=&   \delta_{\nu\mu} a(p,k;q)-q_\nu k_\mu b(p,k;q)  \nonumber \\ 
+ p_\nu q_\mu c(p,k;q)
&+& q_\nu p_\mu d(p,k;q) + p_\nu p_\mu e(p,k;q)) \ , \nonumber \\
\eeq
could be written as follows:

\beq\label{H12}
H_1(q,k) &=& a(-q,k;q-k) - q^2 \left( b(-q,k;q-k) \right. \nonumber \\
&+& \left. d(-q,k;q-k) + 
e(-q,k;q-k) \right) \nonumber \\
H_2(q,k) &=& q^2 \left( b(-q,k;q-k) - c(-q,k;q-k) \right) \ ; \nonumber \\ 
\eeq

We can at this point make our first hypothesis: the scalar factors present in that 
decomposition are regular when one of their arguments 
goes to zero while the others are kept finite. 
Thus, we suppose that 
\beq\label{Hyp1}
a(-r,r-p;p) = a_1(p^2) + {\cal O}(p \cdot r)  \ ,
\eeq
and the same for the other scalars in the particular kinematic configurations 
we shall encounter. We adopt the notations: $a_i(p^2)$, $b_i(p^2)$, 
$c_i(p^2)$ and so on; where the subindex $i$ means that their $i$-th argument 
is a zero momentum. 

The most general tensorial decomposition of the three-gluon vertex, 
$\Gamma_{\lambda \mu \nu}$ (of course, the antisymmetric color tensor $f^{abc}$ is factorised) 
is given in ref. \cite{Ball80}. We will be interested in the limit of one vanishing gluon momentum 
while the two others remain finite. Such a limit deserves a careful analysis in the framework 
of WST identities because of the interplay of gluon and ghost propagator singularities and those 
of scalar functions in the decomposition~\cite{US}.  When one of the momenta is zero the three-gluon 
vertex reduces to  (cf. ref.~\cite{Chetyrkin:2000dq}):
\beq\label{Def3}
&&\Gamma_{\lambda\mu\nu}(q,-q,0) = \nonumber \\
&& \left( 2\delta_{\lambda\mu}q_\nu - \delta_{\lambda\nu}q_\mu - 
\delta_{\nu\mu}q_\lambda\right) T_1(q^2) \\
&-& \left(  \delta_{\lambda\mu} - \frac{q_\lambda q_\mu}{q^2} \right) q_{\nu}T_2(q^2) +
q_\lambda q_\mu q_\nu T_3 (q^2). \nonumber
\eeq
For our purposes here, we will only assume that the limit of one vanishing gluon momentum 
can be safely taken, {\it i.e.}~\footnote{It is shown in~\cite{Ball80}, on a perturbative basis, that the vertex remains finite when 
one takes  the limit $r \to 0$ while keeping the two other momenta fixed. Our hypothesis amounts to assuming that 
this result survives beyond perturbation theory.}:
\beq\label{Hyp2}
\Gamma_{\lambda \mu \nu}(q-r,-q,r) = \Gamma_{\lambda\mu\nu}(q,-q,0) \ + o(1) .
\nonumber \\ 
\eeq

\section{WSTI and IR propagators}
\label{STID}

The Ward-Slavnov-Taylor (\cite{Taylor}) identity for the three-gluon function
reads
\begin{equation}
\label{STid}
\begin{split}
p^\lambda\Gamma_{\lambda \mu \nu} (p, q, r) & =
\frac{F(p^2)}{G(r^2)} (\delta_{\lambda\nu} r^2 - r_\lambda r_\nu) \widetilde{\Gamma}_{\lambda\mu}(r,p;q) 
\\ & -
\frac{F(p^2)}{G(q^2)} (\delta_{\lambda\mu} q^2 - q_\lambda q_\mu) \widetilde{\Gamma}_{\lambda\nu}(q,p;r).
\end{split}
\end{equation}
We shall now study the behaviour when $r \rightarrow 0$ while  keeping $q$ and $p$ finite 
and apply decompositions (\ref{Def2},\ref{Def3}) and the hypotheses (\ref{Hyp1},\ref{Hyp2}) 
to replace the vertices in \eq{STid}. Then, if one only retains the leading terms, STI 
reads 
\beq\label{STid2}
&& T_1(q^2) \left( q_\mu q_\nu - q^2 \delta_{\mu \nu} \right) \ + \ q^2 q_\mu q_\nu \ T_3(q^2) \ + \ o(1) \nonumber \\
&=&\frac{F(q^2)}{G(r^2)} \big[ a_1(q^2) \left( r^2 \delta_{\mu\nu} -r_\mu r_\nu \right) \nonumber  \\
&& + \ b_1(q^2) q_\mu \left( r^2 q_\nu -(q \cdot r) r_\nu\right) + o(r^2) \big] \nonumber \\
&+& \frac{F(q^2)}{G(q^2)} 
\big[ a_3(q^2) \left( q_\mu q_\nu - q^2 \delta_{\mu \nu} \right) + o(1)
\big]  \ .
\eeq
Thus, if one multiplies both l.h.s. and r.h.s of this~\eq{STid2} by $r_\nu$, we obtain:
\beq\label{STlimr0}
&&T_1(q^2)\left(q_\mu \ (q \cdot r) -  q^2 r_\mu \right) \nonumber \\
&+& q^2 q_\mu (q \cdot r) \ T_3(q^2)\ 
+ \ o(r \cdot q) \\
&=& 
\frac{F(q^2)}{G(q^2)} \ a_3(q^2) \ \left( q_\mu (q \cdot r) - q^2 r_\nu \right) \  
+ \ o(r \cdot q) \ ;
\nonumber
\eeq
where the first term of r.h.s. of~\eq{STid2} vanishes because it is transverse to $r_\nu$. 
Thus, by identifying both r.h.s and l.h.s of \eq{STlimr0}, one is led 
to the familiar relations  (\cite{Chetyrkin:2000dq}):
%
\beq
\label{WIhab}
&&T_1(q^2) = \frac{F(q^2)}{G(q^2)}a_3(q^2) \nonumber \\
&&T_3(q^2) = 0 \ .
\eeq
Now, let us multiply both r.h.s and l.h.s. of~\eq{STid2} by $q_\mu$ and apply that $T_3$ has been seen to be 
exactly 0 in \eq{WIhab} and we obtain then:

\beq\label{STlimq0}
&&\frac{F(q^2)}{G(r^2)} \ r^2 \ \big[ \left( a_1(q^2) + q^2 b_1(q^2) \right) \\ 
&\times& \left( q_\nu - \frac{(q \cdot r)}{r^2} \ r_\nu \right) \ + o(1) \big] \ = \ o(1) \ .
\nonumber 
\eeq
Thus,  if $a_1(q^2)\neq 0$ or $b_1 (q^2)\neq 0$ (and, indeed, one knows from  
perturbation theory that at large momenta $a_1$ = 1, cf.~\cite{Chetyrkin:2000dq,Taylor}) 
(\ref{STid}) implies
\beq
\lim_{r \to 0} \frac {G(r^2)}{r^2} \to \infty \ ,
\eeq
or, in other words, that {\bf the gluon propagator diverges in the infrared limit}.  If we stick 
to the commonly accepted idea that G behaves as a power in the infrared 
($G(p^2) \sim (p^2)^{\alpha_G}$) then $\alpha_G < 1$ is to be concluded. 
Another attractive possibility would be to suppose an infrared behaviour 
less divergent than any power as, for instance, that of 
the form : $G(p^2) \sim p^2 \log^\nu(p^2)$ for some 
positive $\nu$.  This will be considered in more detail in a 
forthcoming paper~\cite{US}.

We can also, instead of letting $r \to 0$, study now the behaviour when $p\rightarrow 0$ of 
Eq.(\ref{STid}) as is done in \cite{Chetyrkin:2000dq}.
The dominant part of the l.h.s. of (\ref{STid}) reads:
\beq
(2\delta_{\mu\nu}p.q-p_\mu q_\nu-p_\nu q_\mu) \ a_3(q^2) \frac{F(q^2)}{G(q^2)} \nonumber \\
-(\delta_{\mu\nu}-\frac{q_\mu q_\nu}{q^2}) (p.q) T_2(q^2) \ ;
\eeq
where the results in \eq{WIhab} have been implemented. 
Let us now multiply both sides with $q^\mu$ and keep only the leading tems in $p$ and
one obtains : 
\beq
&&(q_\nu (p\cdot q) - q^2 p_\nu) a_3(q^2) F(q^2) =  \nonumber \\
&&(q_\nu (p\cdot q) - q^2 p_\nu) F(p^2) (a_2(q^2) - q^2 d_2(q^2)) \nonumber \\ 
&&+ {\cal{O}}(p^2)
\eeq
that of course can be true only if $F(p^2) $ goes to some finite limit when $p^2\to 0$ 
and whence, in terms of scalars,
\beq
F(q^2)\,\simquand{q^2\to 0}\, F(0) \ \frac{a_2(q^2) - q^2 d_2(q^2)}{a_3(q^2)}  \ ,
\eeq
where $a_2/a_3 \to 1$ as $q^2 \to 0$~\cite{Chetyrkin:2000dq}.

Let us repeat here that all these considerations are valid only when our regularity hypotheses 
about the ghost-gluon scalar factors and about the three-gluon vertex (see (\ref{Hyp1},\ref{Hyp2})) 
are satisfied. \underline{Under those hypotheses} one obtains important constraints on the gluon 
and ghost propagators - namely that they are divergent in the
zero momentum limit. Let us now briefly analyze the ghost propagator Dyson-Schwinger equation (GPDSE).

\section {Ghost DSE: the case $\alpha_F = 0$}

In a previous paper (the first one of ref.~\cite{Boucaud:2005ce}) we studied all the classes of 
solutions for the GPDSE, that can be pictured, in diagrammatic form, as: 
%
\beq
\left( F^{ab}(q^2) \right)^{-1}
= \left( F_{\rm tree}^{ab}(q^2) \right)^{-1} -
\rule[0cm]{2cm}{0cm}
\nonumber 
\eeq
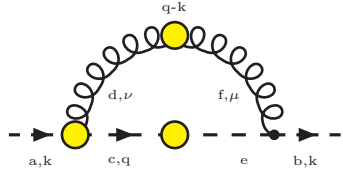
\vspace{\baselineskip}
\begin{tiny}
\beq
\rule[0cm]{1.3cm}{0cm} \ghostSD %
\nonumber  
\eeq\end{tiny}
%

Let us first recall that the { \underline{unsubtracted}} GPDSE is actually meaningless since 
the integral  in its right hand side is UV-divergent, behaving as $\int dq^2 \frac{1}{q^2} 
\left( 1 + 11 \alpha_s/(2 \pi) \log(q/\mu)\right)^{-35/44}$. A way out of this difficulty 
would be to renormalise the  equation  to deal properly with its UV divergencies. 
Instead of that, we preferred to study the following subtracted version of bare  
GPDSE equation for two scales $\lambda k$ and $\kappa \lambda k$ (see Eq.(14) of the first 
paper quoted in ref.~\cite{Boucaud:2005ce})  with $k$ the external ghost momentum and 
$\kappa$ some fixed number ($<1$). $\lambda$ is an extra parameter that we shall ultimately 
let go to $0$ in order to study the infared behaviour of the GPDSE.
This subtracted version of the GPDSE reads (see Eq.(14) of the first paper quoted 
in ref.~\cite{Boucaud:2005ce}):
\beq\label{SDeq}
&&\frac{1}{F(\lambda k)}-\frac{1}{F(\kappa \lambda k)} = \\
&&g_{B}^2 N_c\int \frac{d^4 q}{(2\pi)^4}
\Big(\frac{F(q^2)}{q^2} \left(\frac{(k\cdot q)^2}{k^2}-q^2\right) \nonumber \\
&&\times\Big[\frac{G((q-\lambda k)^2)H1(q,\lambda k)}{(q-\lambda k)^2} -
 (\lambda \to \kappa \lambda)
\Big] \Big) \ , \nonumber
\eeq
where $H_1$ is the particular combination of the scalars defined in \eq{Def2} 
playing the GPSDE game. Furthermore, a proper dimensional analysis of \eq{SDeq} requires to cut the integration 
domain in its r.h.s. into two pieces by introducing some additional scale $q_0^2$ (of the order of $\Lambda_{\rm QCD}^2$). 
Clearly the external momentum is not the only relevant scale in the problem and $\Lambda_{\rm QCD}$, without which 
it would not be understandable that the UV behaviour differs drastically from the IR one, must be 
taken into account. A careful dimensional analysis of the integrals extended over both domains, $q^2 > q_0^2$ and 
$q^2 < q_0^2$, is mandatory~\cite{Boucaud:2005ce}. In the second one --and only there-- we will  initially  use
 the common, convenient, but not really
  justified assumption of a power-law behaviour of the propagators in
  the deep infrared:
 \beq\label{FG}
 F(k^2) &\sim& \left( \frac {k^2}{q_0^2}\right)^{\alpha_F} ,\
G(k^2) \sim \left( \frac {k^2}{q_0^2}\right)^{\alpha_G}.
\eeq
We shall not repeat here the details of our scaling analysis of \eq{SDeq}~\footnote{The analysis done 
in~\cite{Boucaud:2005ce} missed some possible solutions (for instance, the case $\alpha_F=0$, $\alpha_G < 1$) 
mainly because of the fact that we had rejected the 
possibility of non-analytic sub-dominant terms in the dressing functions} 
and simply summarize our conclusions in the following 2 tables.
It is often claimed, after the study of the GPDSE, that $2\alpha_F + \alpha_G=0$. 
In fact, as can be seen in next tab.~\ref{tab1}, this results emerges only~\footnote{The regularity of the 
ghost-gluon vertex is also needed as was discussed in~\cite{Boucaud:2005ce}.} after assuming 
$\alpha_F \neq 0$ and discarding (reasonably) $\alpha_F=-1$.

\begin{table}[h]
\begin{center}
\begin{tabular}{c}
$\alpha_F \neq 0$ \\
\begin{tabular}{|c|c|c|}
\hline
$\alpha_F+\alpha_G$ &r.h.s. & constraint \\
\hline
$ > 1$ &
$\lambda^2$ & 
$\alpha_F=-1$
\\
\cline{1-3}
$=1$ &
$\lambda^2 \log{\lambda}$ & 
excluded
\\
\cline{1-3}
$< 1$ &
$(\lambda^2)^{\alpha_F+\alpha_G}$ & 
$2 \alpha_F+\alpha_G= 0$ \\  
\hline
\end{tabular}
\end{tabular}
\end{center}
\caption{Constraints imposed by the GPDSE to the critical behaviour of ghost and gluon 
propagators for the case $\alpha_F \neq 0$. The second column shows the behaviour on $\lambda$ 
($\lambda \to 0$) of \eq{SDeq}'s r.h.s., while l.h.s. 
behaves as $(\lambda^2)^{-\alpha_F}$.}
\label{tab1}
\end{table}

However, if $\alpha_F=0$ another solutions are also compatible with 
GPDSE (see tab.~\ref{tab2}).

\begin{table}[h]
\begin{center}
\begin{tabular}{c}
$\alpha_F=0$ \\
\begin{tabular}{|c|c|c|}
\hline
$\alpha_F+\alpha_G$ &r.h.s. & constraint \\
\hline
$ > 1$ &
$\lambda^2$ & 
$F(q^2)=A+Bq^2$
\\
\hline
$=1$ &
$\lambda^2 \log{\lambda}$ & 
 $F(q^2)=A+Bq^2 \log{q^2}$
\\
\cline{1-3}
$< 1$ &
$(\lambda^2)^{\alpha_G}$ & 
$F(q^2) = A+Bq^{2 \alpha_G}$\\  
\hline
\end{tabular}
\end{tabular}
\end{center}
\caption{The same constraints analysed in tab.~\ref{tab1} but here for $\alpha_F=0$. 
The l.h.s. of \eq{SDeq} behaves now as the next-to-leading 
term of the deep infrared expansion of $F(q^2)$ (third column).}
\label{tab2}
\end{table}

Some recent lattice results seem to exclude the standard 
($2 \alpha_G + \alpha_F=0$)-solution~\cite{Boucaud:2005ce,Sternbeck:2005re}.
If one admits these results (lattice also discards $\alpha_F=-1$), then one is led to 
conclude that GPDSE implies $\alpha_F=0$. 

Furthermore, it was shown in ref.~\cite{Boucaud:2006if} that the r.h.s. of \eq{SDeq} is the sum of 
two terms behaving respectively as
$\lambda^{2{\rm Min}(\alpha_F+\alpha_G+\alpha_\Gamma,1)}$ and $\lambda^2$ when 
$\lambda\rightarrow 0$.
So it behaves as $\lambda^2$ when $\alpha_F=0$. Then, one can proves 
that for any $\kappa$ there is a value of $\lambda$ and $c$ such that 
\beq\label{bound}
\vert \frac{1}{F(\lambda k)}-\frac{1}{F(\kappa^n \lambda k)}\vert \leq c
 \frac{1-\kappa^{2n}}{1-\kappa^2}\lambda^2.
\eeq
So $F\rightarrow\infty$ when $\lambda\rightarrow 0$ is excluded because taking the limit of the above
expression when $n\rightarrow\infty$ we should have $\vert \frac{1}{F(\lambda k)}\vert
\leq c \frac{1}{1-\kappa^2}\lambda^2$ and F would diverge as or more rapidly than 
$\frac{1}{\lambda^2}$ implying $\alpha_F \leq -1$ in contradiction with the hypothesis 
$\alpha_F=0$.
Let us remark that $F\rightarrow 0$ is also excluded: Eq. (\ref{bound}) implies 
$\vert \frac{1}{F(\kappa^n \lambda k)}\vert \leq \vert \frac{1}{F(\lambda k)}
\vert+c\frac{1-\kappa^{2n}}{1-\kappa^2}\lambda^2$ and $\frac{1}{F(\kappa^n \lambda k)}$ 
cannot tend to  infinity when $n\rightarrow\infty$. It should be emphasized that the dimensional 
analysis driving to \eq{bound} is also valid if $F(q^2)$ is admitted to behave in a way other 
than a power. Thus, \textbf{if a leading power behaviour is discarded for the ghost dressing function, it has 
to be finite and $\neq 0$ in the IR limit}. 


\begin{figure}
\begin{center}
\resizebox{0.30\textwidth}{!}{%
  \includegraphics{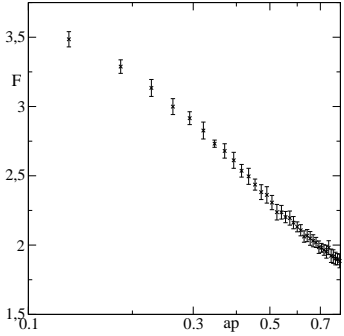}
}
\end{center}
\caption{F(p) from a SU(2) simulations on a $48^4$ lattice at $\beta=2.3$}
\label{Fig1}       
\end{figure}
%
%
\begin{figure}
\begin{center}
\resizebox{0.35\textwidth}{!}{%
  \includegraphics{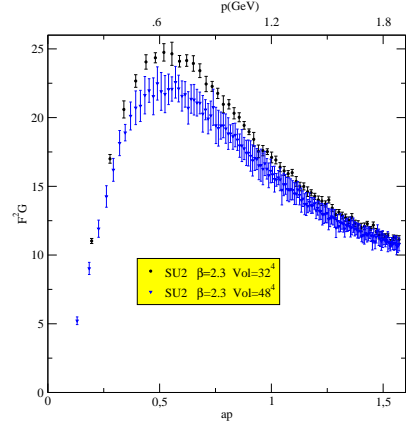}
}
\end{center}
\caption{$F^2 G$ from lattice simulation for $SU(2)$ ($32^4 $ and $48^4,\beta_{SU(2)}=2.3$) gauge groups. 
$2\alpha_F + \alpha_G = 0$ implies a constant in the infrared domain.}
\label{Fig2}       
\end{figure}
%
%

\section{conclusion}

We derive from the Ward-Slavnov-Taylor identity for the three-gluon and ghost-gluon vertices, 
after assuming their regularity that gluon propagator diverges and ghost dressing function remains finite as the momentum goes 
to zero. A dimensional analysis of the GPDSE, provided that we trust the lattice 
results excluding $2 \alpha_F + \alpha_G = 0$~\cite{Boucaud:2005ce,Sternbeck:2005re} and $\alpha_F = -1$, leads to conclude 
independently that the ghost dressing function remains finite at zero momentum~\cite{Boucaud:2006if} 
(see tabs.~\ref{tab1},\ref{tab2}). Both GPDSE and WSTI constraints will offer compatible solutions provided that one admits 
non-analytic sub-leading terms for the low momentum expansion of dressing functions. 

On the other hand, such a solution respecting WSTI and GPDSE constraints still match in the 
present picture of lattice knowledge about the IR behavior of propagators and vertices. 
The current simulations of ghost-gluon vertex seems to discard $2 \alpha_F + \alpha_G=0$ but 
those of ghost and gluon propagators cannot yet exclude or confirm the smooth divergences
we propose as a way out~\cite{Sternbeck:2005re,Adelaide,Becirevic:1999hj} (as an example, see Fig.~\ref{Fig1} from 
ref.~\cite{Boucaud:2006if} or Fig.~\ref{Fig2} from ref.~\cite{Boucaud:2005ce}). 
A non-power behaviour (logarithmic, for instance) could be specially elusive for 
lattice extrapolations at infinite volume. Of course, new simulation results on bigger lattice volumes (or with twisted 
boundary conditions~\cite{Tok:2005ef}) and careful extrapolations will be very welcome to dig into this 
matter.

This is a very interesting task to be acomplished, because either such a logarithmic (or similar) behaviour is found 
or one is led to conclude that the tensorial decomposition of ghost-gluon or three-gluon 
vertex admits non-regularities. 

 
%

\end{document}